\documentstyle[bo99,epsfig]{article}
\def\ltsima{$\; \buildrel < \over \sim \;$}
\def\lsim{\lower.5ex\hbox{\ltsima}}
\def\gtsima{$\; \buildrel > \over \sim \;$}
\def\gsim{\lower.5ex\hbox{\gtsima}}

\def\mdot {\dot M}

\def\cmdue {\rm \ cm^{-2}}

\newcommand{\ergs}{\rm \ erg \; s^{-1}}
\def\msole {~M_{\odot}}
\newcommand{\be}{\begin{equation}}
\newcommand{\en}{\end{equation}}
\def\refb{\par\noindent\hangindent 15pt}

\title{X--ray transients in quiescence}
\author{Sergio Campana}
\affil{Osservatorio astronomico di Brera, Via E. Bianchi 46,
Merate (LC), I-23807 Italy; e-mail: campana@merate.mi.astro.it}

\begin{document}

\maketitle

\begin{abstract}
Transient X--ray binaries remain in their quiescent state for a long time (months to 
hundred years) and then bright up as the most powerful sources of the X--ray sky.
While it is clear that, when in outbursts, transient binaries are powered 
by accretion, the origin of the low luminosity X--ray emission that has 
been detected in the quiescent state has different interpretations
and provides the unique opportunity for testing different accretion regimes.
In this paper we concentrate on the various aspects of the accretion physics at 
low rates onto compact objects. 
We describe the observational panorama of quiescent emission for the
three classes of X--ray transients and try to interpret these data in light of the 
different regimes accessible at such low mass inflow rates.
\keywords{X--ray: stars --  Accretion, accretion disks -- Black
hole physics -- Stars: neutron}
\end{abstract}

\section{Introduction}

X--ray transients are classified based on their outburst spectral
properties (e.g. White et al. 1984): hard X--ray transients 
(HXRTs), soft X--ray transients (SXRTs) and ultra soft X--ray transients (USXRTs).
This classification is successful since it reflects the true nature of these 
systems: HXRTs host a high magnetic field neutron 
star (NS) accreting from a high mass, usually Be, companion star$\,$\footnote{Here 
I consider only sources containing fastly spinning NS ($P\lsim 10$ s), which 
share several properties with SXRTs.} (for a review see Bildsten et al. 1997);
SXRTs host a low field NS accreting from a late type, usually K-M, star (e.g. 
Campana et al. 1998a) and USXRTs consisting of a black hole candidates (BHCs) 
accreting from a low mass companion too (e.g. Tanaka \& Shibazaki 1996).
Transient binaries are characterised by an X--ray luminosity that 
varies over many orders of magnitude, allowing to probe different 
conditions and accretion regimes that are unaccessible to persistent 
(bright) sources. 

In this paper I will first review the observational properties of transient X--ray 
sources and then challenge these with simple theoretical models.

\section{Black hole transients}

BHCs are usually faint in their X--ray quiescent state and strong upper limits 
exist for a number of them (Menou et al. 1999; Campana \& Stella 2000).
The first, and only short orbital period (7.8 hr), BHC detected in quiescence 
to date is A~0620--00. This is the prototype BHC: after a very bright outburst 
peaking at $\sim 7$ Crab, the source slowly returned to quiescence and only 
ROSAT, years later, was able to reveal it. 
The 0.1--2.4 keV luminosity is $\sim 6\times 
10^{30}\ergs$ (for a distance of 1.2 kpc and by fixing the column density to 
$N_H=1.2\times 10^{21}\cmdue$; McClintock et al. 1995). Due to the small number 
of photons ($\sim 40$) however, the spectrum is very poor and can be well fit 
by a variety of single component models. In particular, it cannot be excluded 
that such a low luminosity arises from the K dwarf companion (see also Bildsten 
\& Rutledge 2000).

The other two BHCs detected in quiescence are characterised by longer orbital 
periods ($>2.5$ d) and therefore higher average mass inflow rates, based on 
evolutionary models. GS 2023+338 (V~404 Cyg) was detected 
at $L\sim 2\times 10^{33}\ergs$ with ASCA and BeppoSAX (Narayan et al. 1997; 
Campana et al. 2000c). The spectrum is well fit by either a power law (photon 
index $\Gamma\sim 1.5-2$) or a bremsstrahlung ($k\,T_{\rm br}\sim 5-10$ keV).
GRO J1655--40 was detected at $L \sim 2\times 10^{32}\ergs$ (Hameury 
et al. 1997), with a spectrum that can be described by a power law model with 
$\Gamma\sim 1.5$.

\subsection{ADAF}

The very low luminosity of BHCs in quiescence has stimulated a number of works.
The paradigm is now represented by advection-dominated accretion flow (ADAF) models,
where the radiative efficiency is very low ($\sim 10^{-4}-10^{-3}$) and most 
of the gravitational energy of the inflowing matter is stored as thermal and/or 
bulk kinetic energy and advected towards the collapsed star (e.g. Narayan et 
al. 1996; Narayan et al. 1997; Menou et al. 1999). Solutions of this type exist for 
sub-Eddington mass accretion rates ($\mdot < 0.1-0.01\ \mdot_{\rm Edd}$).
In this regime the bolometric luminosity scales approximately as $\dot M^2$ (as 
opposed to the $\dot M$ scaling of standard accretion; cf. Fig. 1). 
These models (which where modified under way to include the optical/UV 
luminosity as produced by the ADAF itself; cf. Narayan et al. 1997) provide good 
fit to the multi-wavelength spectra of quiescent BHCs. 

\begin{figure*}[!htb]
\psfig{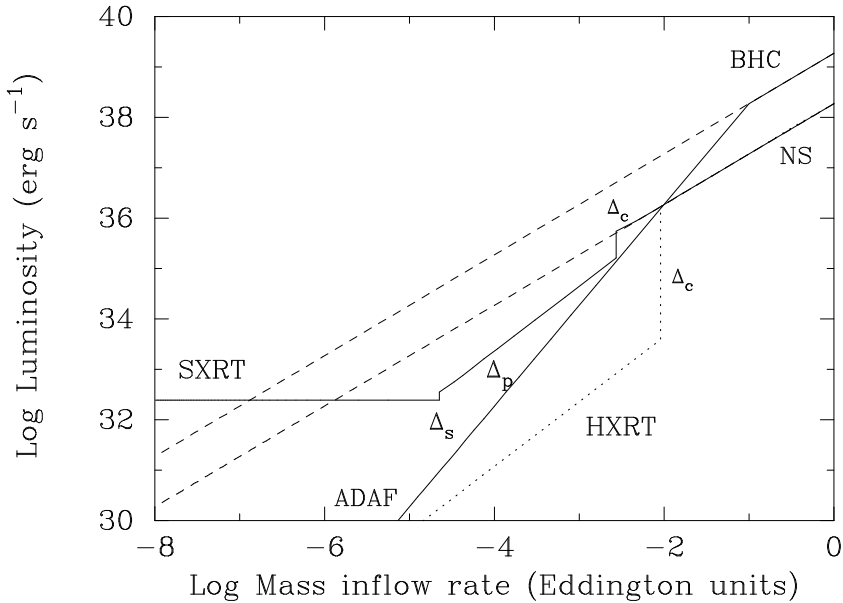}
\caption[h]{Luminosity versus the mass inflow rate (in Eddington units
$\mdot \sim 1.4\times 10^{18}\,M/\msole$ g s$^{-1}$)  
for different accretion regimes onto BHs and NSs.
The upper line refers to a $14\msole$ BH. 
The dashed line marks the standard accretion regime  
(efficiency $\epsilon = 0.1$) and 
the continuous line the ADAF model. The lower lines refer to accretion 
onto a $1.4\msole$ NS. The continuous line gives the luminosity produced by 
a 2.5 ms spinning, $B=10^8$ G NS in different regimes, the dotted line
by a 4 s, $B=10^{12}$ G NS. The lower dashed line refers to accretion onto the NS 
surface.}
\end{figure*}

\section{Hard X--ray Transients}

\begin{figure*}[!htb]
\centerline{\psfig{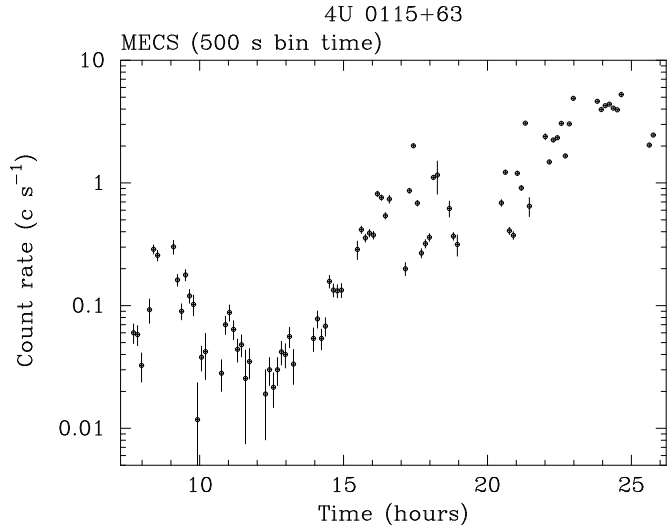}}
\caption[h]{BeppoSAX MECS light curve of 4U 0115+63. The orbital phase 
is just before periastron, based on the ephemerides of Cominsky et al. (1992).}
\end{figure*}

Despite a considerable increase in the number of new HXRTs discovered in the 
last few years thanks to RXTE, observational data on the transition to 
quiescence of HXRTs are still sparse.
For only a few systems (V 0332+53 Stella et al. 1986; 4U 0115+63 Tamura et al. 
1992) there were indications of a sudden turn off of the X--ray luminosity 
when the sources achieve a level of $\sim 10^{36}\ergs$. 

Even more rare are the observations of HXRTs in quiescence.
In the last few years, only the HXRT A 0538--66 in the Large Magellanic Cloud 
(containing the fastest accreting NS at 69 ms) has been positively 
detected in quiescence.
During the ROSAT all-sky survey two weak outbursts were
detected, with peak luminosities of $\sim 4$ and $\sim
2\times10^{37}\ergs$ in the 0.1--2.4 keV range (Mavromatakis \& Haberl
1993) and a similar weak outburst was detected by ASCA at $\sim 6 \times 
10^{36}\ergs$ (1--10 keV; Corbet et al. 1997; Corbet 1996).
A 0538--66 was detected several times at a level of about 
$10^{34}-10^{35}\ergs$ during ROSAT PSPC serendipitous pointings (Campana 1997).
ASCA and ROSAT observations gave a first indication of the spectrum 
at such low rates: the ASCA spectrum is well fit by a power law (photon index
$\Gamma\sim 2$) plus a soft component, e.g. a black body with temperature
$\sim 3$ keV and equivalent radius of $\sim 2$ km. The ROSAT PSPC spectrum 
at a factor of 10--100 lower luminosity can be fit by a black body model
with much smaller temperature ($\sim 0.2$ keV) and larger radii ($\sim 70$ km).
The presence of a hard power law however cannot be excluded.
Recently, we obtained BeppoSAX observations of three fast spinning HXRTs 
(A 0538--66, V 0332+53 and 4U 0115+63) during their quiescent states
(Gastaldello et al. 2000). 
The most striking results comes from the observation of 4U 0115+63. 
A 15 hr BeppoSAX observation shows a variation in the count 
rate by a factor of $\sim 250$ (cf. Fig. 2; Campana et al. 2000a). 
A time-resolved spectral analysis reveals that this variation is 
intrinsic to the source, which does not change its spectrum (hard power 
law with photon index $\Gamma\sim 1$) nor its column density (a few 
$10^{22}\cmdue$, washing out any soft component).
The mean 0.1--200 keV luminosity in each interval varies from 
$\sim 2\times 10^{34}\ergs$ to $\sim 2\times 10^{36}\ergs$ (at 4 kpc).
Pulsations were detected all the way down to the smaller fluxes.

\subsection{Propeller regime}

In the relatively slow ($P\geq 1$~s) and high magnetic field ($B\sim 10^{12}$~G) 
NSs of HXRTs accretion onto the surface can 
take place as long as the magnetospheric radius ($r_{\rm m}$, at which the NS 
magnetic field starts dominating the motion of the infall matter) is smaller
than the corotation radius ($r_{\rm cor}$, at which matter in Keplerian 
rotation orbits at the same angular frequency of the NS).
In this regime, the accretion luminosity is given by $L(R)=G\,M\,
\mdot/R$. 
As the mass inflow rate decreases, $r_{\rm m}$ expands till it reaches the 
corotation radius. For smaller rates, matter getting attached to the NS field
lines at  $r_{\rm m}$ experiments a centrifugal force stronger than gravity and 
gets expelled. The source starts to be centrifugally inhibited (propeller regime) 
at a luminosity of \footnote{We use here simple spherical accretion theory. This 
is a reasonably accurate approximation when the accretion
disk at the magnetospheric boundary is dominated by gas pressure. For a more general
approach see e.g. Campana et al. (1998a).}
\begin{eqnarray}
L_{\rm cb}&\simeq& 2 \times 10^{36}\, B_{12}^2\, M_{1.4}^{-2/3}\, R_6^5\, 
P_{4\,{\rm s}}^{-7/3}\ {\rm erg\ s}^{-1} \nonumber \\ 
&\simeq& 5 \times 10^{35}\, B_8^2\, M_{1.4}^{-2/3}\, R_6^5\, 
P_{2.5\,{\rm ms}}^{-7/3}\ {\rm erg\ s}^{-1}  \ 
\label{lmin}
\end{eqnarray}
($B=10^{12}\,B_{12}$ G -- $B=10^8\,B_8$ G, $P=4\, P_{\rm 4\, s}$~s -- $P=2.5\times 
10^{-3}\, P_{\rm 2.5\, ms}$~s, $R=10^6\, R_6 
$~cm and $M= 1.4\, M_{1.4}\msole$ are the magnetic field, spin period, radius 
and mass of the NS, respectively).
Matter, being stopped at $r_{\rm m}$ rather than $R$, releases a lower 
accretion luminosity. The accretion luminosity gap, $\Delta_{\rm c}$, 
across the centrifugal barrier is (Corbet 1996; Campana \& Stella 2000)
\begin{eqnarray}
\Delta_{\rm c} = {{r_{\rm cor}} \over R} = \Bigl( {{G\,M\,P^2}\over 
{4\,\pi^2\,R^3}} \Bigr)^{1/3} 
& \simeq & 620 \, P_{\rm 4\,s}^{2/3}\,M_{1.4}^{1/3}\,R_6^{-1} \nonumber \\
& \simeq & 3\, P_{\rm 2.\,5 ms}^{2/3}\,M_{1.4}^{1/3}\,R_6^{-1}
\ . 
\label{gapa}
\end{eqnarray}
$\Delta_{\rm c}$ depends almost exclusively on the
spin period $P$ and is basically a measure of how deep $r_{\rm cor}$ is 
in the potential well of the NS. 

This simple picture is challenged by the recent observations of 4U 0115+63.
The centrifugal gap that has been modeled as a step-like transition from the 
accretion to the propeller (or viceversa) regimes has likely been observed.
Assuming a typical mass inflow rate variation as derived from BeppoSAX
observations of 4U 0115+63 in outburst, one can derive a relation between
the observed luminosity and the mass accretion rate $L\propto \mdot^{\alpha}$
which in the case of standard accretion implies $\alpha=1$ and in the propeller
$\alpha=9/7$. We derive a value of $\alpha\sim 30$, indicating that a very small
variation in $\mdot$ induces a huge variation in luminosity.
As a confirmation of this, disk and wind accretion model for a system like 
4U 0115+63 allows for a mass inflow rate variation by a factor of a few in 15 hr, 
at most (e.g. Raguzova \& Lipunov 1998). 

At variance with simple model predictions, X--ray pulsations are detected all 
the way down the lower part of the gap (even if with decreasing pulsed fractions). 
The most straightforward interpretation is that some matter still leaks the 
centrifugal barrier likely from the highest magnetic latitudes.
We conclude that we are observing for the first time the transition 
from the propeller regime to the standard accretion onto the NS surface. 
This transition is very fast and opens the possibility 
to study in detail how the centrifugal mechanism works (see also Campana et al. 
2000a, Pizzolato et al. 2000).

\section{Soft X--ray Transients}

SXRTs in quiescence were the target of early X--ray astronomy 
missions such as Einstein (Petro et al. 1981) and EXOSAT (van Paradijs 
et al. 1987). These data however provide very poor spectral information. 
ROSAT gave for the first time a clear assessment in the field, 
revealing the SXRT prototype source Aql X-1 on three occasions at a 
level of $\sim 10^{33}\ergs$ (0.4--2.4 keV) with a very soft spectrum 
(e.g. a black body with a temperature of $k\,T_{\rm bb}\sim 0.3$ keV 
and an equivalent radius of $\sim 10^{5}$ cm; Verbunt et al. 1994).

The number of SXRTs detected in quiescence is now increasing thanks to 
BeppoSAX and ASCA: Aql X-1 (Campana et al. 1998b; Asai et al. 1998), 
Cen X-4 (Asai et al. 1996; Campana 
et al. 2000b), 4U~1608--522 (Asai et al. 1998), 4U 2129+47 and EXO 0748--676 
(Garcia \& Callanan  1999) and most recently SAX J1808.4--3658 (Stella et 
al. 2000) and X 1732--304 (Guainazzi et al. 1999). All these sources
have X--ray luminosities in the $10^{32}-10^{33}\ergs$ range (e.g. Campana et
al. 1998a). Together with the soft spectral component which
is present in all SXRTs in quiescence observed to date (usually modeled as 
a black body of $k\,T_{\rm bb}\sim 0.1-0.3$ keV), 
a hard power law can be revealed in the best studied sources (Aql X-1, Cen X-4 and 
X 1732--304). This power law (with photon index $\sim 1.5-2$) makes up 
$\sim 50\%$ of the 0.5--10 keV luminosity. This two components spectrum is 
becoming the `canonical' spectrum for SXRTs in quiescence and further 
confirmations will come from Chandra and XMM-Newton pointings.

\subsection{Shock emission}

It had long been suspected that the NSs of persistent and 
transient LMXRBs have been spun up to very short rotation  
periods by accretion torques however, conclusive evidence 
has been been gathered only recently. The best example is 
SAX~J1808.4--3658, a bursting transient 
source discovered with BeppoSAX in 1996 (in't Zand et al. 1998). 
In April 1998, RossiXTE observations revealed a coherent 
$\sim 401$~Hz modulation, testifying to the presence of magnetic 
polar cap accretion onto a fast rotating magnetic NS (Wijnands \& 
van der Klis 1998; Chakrabarty \& Morgan 1998).
Millisecond rotation periods have also been inferred for 7 other 
LMXRBs of the Atoll (or suspected members of the) group through 
the oscillations that are present for a few seconds during type 
I X--ray bursts (for a review see van der Klis 2000). 
Regarding the NS magnetic field, Psaltis \& Chakrabarty (1999) estimate  
for SAX~J1808.4--3658 a value of $B\sim 10^8-10^9$ G, by adopting 
different models for the disk-magnetosphere interaction. Indirect 
evidence for fields of $B\sim 10^8$ G derives also from the 
steepening in the X--ray light curve decay and marked change of 
the X--ray spectrum when the luminosity reaches a level of $\sim 
10^{36}\ergs$ in Aql X-1 (Campana et al. 1998b; Zhang et al. 1998) 
SAX J1808.4--3658 (Gilfanov et al. 1998), XTE 2123--058 (Tomsick et al. 1999) and
Rapid Burster (Masetti et al. 2000), once these changes are 
interpreted in terms of the onset of the centrifugal barrier. 
The spin period and the inferred magnetic field strength of SAX~J1808.4-3658
provide the first direct evidence for the long suspected Low Mass X--Ray Binaries
millisecond pulsars connection. 

As the mass inflow rate decreases further the magnetosphere expands 
until the light cylinder radius, $r_{\rm lc}=c\,P/2\,\pi$, is reached; 
beyond this point the radio pulsar dipole radiation will turn on and 
begin pushing outward the inflowing matter, due 
to a flatter radial dependence of its pressure compared to that of 
disk or radial accretion inflows 
(Illarionov \& Sunyaev 1975; Stella et al. 1994; Campana et al. 1995). 
The equality $r_{\rm m}=r_{\rm lc}$ defines the lowest mass 
inflow rate 
(and therefore accretion luminosity) in the propeller regime. 
An accretion luminosity ratio of 
\be
\Delta_{\rm p}=\Bigl( {{r_{\rm lc}}\over {r_{\rm cor}}} \Bigr)^{9/2} \simeq
440\, P_{\rm 2.5\, ms}^{3/2}\,M_{1.4}^{-3/2}
\label{gapp}
\en
characterises the range over which the propeller regime applies. 
Note that also this ratio depends mainly on the spin period $P$.

Once in the rotation powered regime, a fraction $\eta$ 
of the spin down luminosity, $L_{\rm sd}$, converts to shock emission 
in the interaction between the relativistic wind of the 
NS and the companion's matter flowing through the Lagrangian point. 
Theoretical models indicate that the material lost by the
companion star may take somewhat different shapes, ranging from a bow shock 
to an irregular annular region in the Roche lobe of the NS,  
depending on radio pulsar wind properties and 
the rate and angular momentum of the mass loss from the companion star
(Tavani \& Brookshaw 1993).
$\eta$ may be as large as 0.1 (Tavani 1991) and 
the shock luminosity can be expressed as 
$L_{\rm shock}=\eta\,L_{\rm sd}\sim 2\times 10^{32}\,\eta_{-1}\,B^2_8
\,P_{2.5\,{\rm ms}}^{-4}\ergs$ ($\eta\sim \eta_{-1}\,0.1$). 
The luminosity ratio across the transition from the propeller to the 
rotation powered regime can be approximated as (Stella et al. 1994;
Campana et al. 1998a)
\be
\Delta_{\rm s}={ 3 \over{2\,\sqrt{2}\,\eta}} \, \Bigl( {{r_{\rm g}}
\over {r_{\rm lc}}}\Bigr)^{1/2}\simeq 2\,\eta_{-1}^{-1}\,
P_{\rm 2.5 \, ms}^{-1/2}\,M_{1.4}^{1/2} \ ,
\label{gaps}
\en
\noindent where $r_{\rm g}=G\,M/c^2$ is the gravitational radius.
The energy spectrum due to shock emission is expected to be a power law with 
photon index of $\sim 1.5-2$ that extends from a $\sim 10$ eV to $\sim 100$ 
keV, with both energy boundaries shifting as $B_8\,P_{\rm 2.5 ms}^{-3}$ 
(Tavani \& Arons 1997; Campana et al. 1998a).

Observationally, 
the nature of the two spectral components observed during the quiescent state 
of SXRTs is a matter of debate. 
One possibility is that some matter leaks through the centrifugal barrier
accreting onto the NS surface and produces in turn the observed soft 
component, whereas the hard component arises in an ADAF (Zhang et al. 1998; 
Menou et al 1999). However a clear assessment of this spectral model has not 
been carried out yet, nor self-consistent ADAF models for NSs exist. 
The other possibility complains a 
cooling NS$\,$\footnote{
Concerning the soft X--ray thermal-like component, we note that the effective 
black body radii inferred from spectral fitting are substantially smaller 
than the NS radius. At the same time, radiative transfer calculations for the NS 
atmosphere indicate that the emergent thermal-like X--ray spectrum is 
complex and simple black body fits are likely to underestimate the 
effective emission radius and overestimate the temperature 
by a factor of 3--10 and 2--3, respectively (Rutledge et al. 1999; Brown et 
al. 1998). 
Consequently, it cannot be ruled out yet that thermal emission 
from the whole NS surface powers the soft X--ray component of 
quiescent NS SXRTs.}
(which  contributes to the soft component) working as a radio pulsar, 
the relativistic wind of which generates a shock power law spectrum  
(Campana et al. 1998a). 
This is, at least 
qualitatively, in agreement with the hard power law like X--ray component 
observed in the quiescent X--ray spectrum of Cen~X-4, Aql X-1 and X~1732--304.
The extended power law spectrum expected from shock emission
is also in agreement with the recently determined residual UV spectrum 
of Cen~X-4, which shows no evidence for a turnover down to lowest measured 
UV energies ($\sim 7.5$~eV;  McClintock \& Remillard 2000) and matches quite 
well the extrapolation of the (power law) X--ray spectrum.

\section {Conclusions}

In the last few years a large wealth of new data on quiescent
transient sources has been obtained thanks to BeppoSAX and ASCA.
These observations give us the opportunity to study in some details the
different regimes that can be expected at such low X--ray fluxes, such as
the ADAF models for BHCs, the propeller regime for HXRTs and the shock emission
for SXRTs. 
New and better data will be collected in the next few years thanks to Chandra and 
XMM-Newton, answering to some basic questions that are still open.

Here I list some topics that can be addressed in the next years:

\begin{itemize}

\item{{\bf Quiescent emission of BHCs (1).} ADAF models are now very popular in 
explaining the BHC quiescent emission. However these models face problems
since their optical/UV luminosity (after subtraction of the contribution 
from the companion star) dominates by a factor of $\gsim 10$ over the X--ray 
luminosity, at variance with SXRTs in which the two luminosities are comparable,
at the most.
Based on the latest ADAF models, this optical/UV luminosity
should be ascribed to the ADAF itself, arising from synchrotron processes.
These results indicate that the luminosity swing from the outburst peak to
quiescence, which has been claimed to be different in BHCs and SXRTs (Narayan, 
Garcia \& McClintock 1997), 
does not provide evidence for a significant difference between the two classes.
Therefore, one of the key motivations for considering ADAF models, 
i.e. that most of the energy is hidden beyond the BH event horizon, is considerably 
weakened (see Campana \& Stella 2000).}

\item{{\bf Quiescent emission of BHCs (2).} The only short
orbital period BHC detected in quiescence is A 0620--00 at a level of 
$\sim 10^{31}\ergs$ (0.5--10 keV, e.g. Menou et al. 1999). 
In analogy with the X--ray emission from K stars in RS CVn type binaries 
(which emit up to $\sim 10^{32}\ergs$ in the ROSAT band; Dempsey
et al. 1993) one may argue that, in short orbital period BHC systems, such as 
A 0620--00, the low level X--ray quiescent luminosity 
($\sim 10^{31}-10^{32}\ergs$) might also arise from due to coronal activity of 
the companion star (see also Bildsten \& Rutledge 2000).}

\item{{\bf Centrifugal gap in HXRTs.} BeppoSAX observations of 4U 0115+63 provide
for the first time the opportunity to study the transition from the accretion
regime to the propeller state. Good data have been obtained and their
potential is being exploited, with the aim of understanding
how and when the centrifugal barrier closes (and/or opens; Campana et al. 
2000a; Gastaldello et al. 2000). 
These data also demonstrate that the monitoring of HXRTs in quiescence, 
especially near periastron, is worthy.}

\item{{\bf Quiescent emission of SXRTs.} Two facing models have been proposed
to explain the quiescent emission of SXRTs: one is based on ADAFs and included the 
leaking of matter through the centrifugal barrier to explain the soft X--ray
component; the other one envisages the presence of an active radio pulsar.
One of the most straightforward prediction of the ADAF scenario is that pulsations 
should be detected in the soft component.
In the case of an active radio pulsar, a radio signal might be detected if 
the circumstellar material does not absorb completely the signal.
Doppler maps can also be useful to pinpoint unusual geometries not related
to an accretion disk but to a shock front.} 

\end{itemize}

\begin{acknowledgements}
I thank useful discussions with a number of people, together with part of this 
work has been carried out. In particular, I would like to thank L. Stella, 
S. Mereghetti,  M. Colpi, G.L. Israel, M. Tavani, F. Gastaldello, D. Ricci, 
T. Belloni, D. Lazzati, G. Tagliaferri, S. Covino, G. Ghisellini. 
This work was partially supported through ASI grant.
\end{acknowledgements}

\end{document}